# Applying Empirical Software Engineering to Software Architecture: Challenges and Lessons Learned

Davide Falessi[1], Muhammad Ali Babar[2], Giovanni Cantone[1], Philippe Kruchten[3]

[1] University of Rome "Tor Vergata", DISP, Rome, Italy
[2] IT University of Copenhagen, Denmark
[3] University of British Columbia, ECE, Vancouver, Canada

**Abstract**. In the last 15 years, software architecture has emerged as an important software engineering field for managing the development and maintenance of large, software-intensive systems. Software architecture community has developed numerous methods, techniques, and tools to support the architecture process (analysis, design, and review). Historically, most advances in software architecture have been driven by talented people and industrial experience, but there is now a growing need to systematically gather empirical evidence about the advantages or otherwise of tools and methods rather than just rely on promotional anecdotes or rhetoric. The aim of this paper is to promote and facilitate the application of the empirical paradigm to software architecture. To this end, we describe the challenges and lessons learned when assessing software architecture research that used controlled experiments, replications, expert opinion, systematic literature reviews, observational studies, and surveys. Our research will support the emergence of a body of knowledge consisting of the more widely-accepted and well-formed software architecture. theories.

**Keywords:** *Software architecture, Empirical software engineering.*

# 1 Introduction

### 1.1 Context

One of the objectives of Empirical Software Engineering is to gather and utilize evidence to advance software engineering methods, processes, techniques, and tools (hereafter called *technologies*). According to Basili (1996): "like physics, medicine, manufacturing, and many other disciplines, software engineering requires the same high level approach for evolving the knowledge of the discipline; the cycle of model building, experimentation, and learning. We cannot rely solely on observation followed by logical thought." One of the main reasons for carrying out empirical research is the opportunity to get objective measures (e.g., in the form of statistically significant results) regarding the performance of a particular software development technology (Wohlin et al. 2000). Several researchers have highlighted the need for, and the importance of empiricism in software engineering (Basili et al. 1986; Juristo and Moreno 2006; Kitchenham et al. 2004; Perry et al. 2000). Others have highlighted problems due to a lack of validated data in major software engineering publications (Zelkowitz and Wallace 1998). During the last two decades, empirical software engineering research has achieved considerable results in building our knowledge (Jeffery and Scott 2002), and this in turn has driven important advances in different areas of software engineering. For instance, the application of empiricism has provided solid results in the area of software economics (Boehm 1981) and of value-based software engineering (Biffl et al. 2005). The application of empiricism has also helped improve defects detection techniques (Shull et al. 2006; Vegas and Basili 2005).

During the same period, software architecture has emerged as an important field of software engineering for managing the development and maintenance of large, software-intensive systems. The software architecture community has developed numerous technologies to support the architecture process (analysis, design, and review). Historically, most advances in software architecture have been driven by talented people and industrial experience, but there is now a growing need to systematically gather empirical evidence about the advantages or otherwise of tools and methods rather than just rely on promotional anecdotes or rhetoric (Dybå et al. 2005; Oates 2003). Hence, there is a need to systematically gather and disseminate evidence that will help: i) researchers in their assessment of current research and help them to identify the most promising areas of research, and ii) practitioners in making informed decisions in order to select suitable technologies for supporting the software architecture process (analysis, design, and review).

**1.2 Vision**

We show in Figure 1 the relationships among software architecture theory, empirical theory, empirical assessments, challenges, lessons learned, and empirical results. When researchers attempt to empirically assess software architecture theory, they face challenges from both empirical theory and software architecture theory (see Section 3.2). Empirical theory provides the means to gather and disseminate evidence in order to support the claims of efficiency or efficacy of a particular technology. Software architecture theory provides the hypothesis which will be accepted or rejected. Empirical research can provide the results on which to build and/or assess the theoretical foundations underpinning various software architecture-related technologies (Sjøberg et al. 2008). Experiences and lessons learned from empirically assessing software architecture research represent a valuable — but often underestimated — means of improving the application of the empirical paradigm to software architecture research and practice (see Section 3.3).

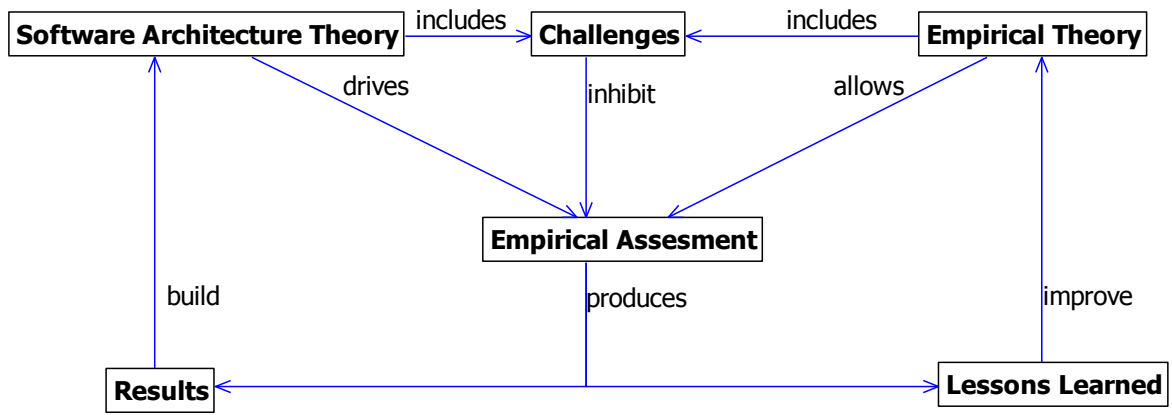

Fig. 1: Relationships between empirical theory and software architecture theory.

## 1.3 Our research

The purpose of this research is to promote and facilitate the application of the empirical paradigm to software architecture. To this end, we present and discuss our experiences, the lessons learned and the challenges faced while applying various empirical research methods (e.g., controlled experiments, replicas, expert opinion, systematic literature review, observation study, and surveys) to assess software architecture technologies. We hope that this work will encourage researchers to carry out high quality empirical studies to evaluate software architecture technologies. In particular, researchers can exploit the lessons we have learned and will understand the challenges. Additionally, we highlight the importance of greater interaction between the empirical software engineering and software architecture communities.

The novelty of this paper, which is an extension of (Falessi et al. 2007), lies in the characterization of the empirical paradigm with respect to its applicability to software architecture. Therefore, the content of this paper should be considered as a complement to, and a specialization of, past general empirical software engineering work (Basili 1996; Juristo and Moreno 2006; Kitchenham 1996; Sjøberg et al. 2007; Wohlin et al. 2000; Zelkowitz and Wallace 1998) .

### 1.4 Structure

The rest of the paper is organized as follows: Section 2 presents the motivation and background for this research. Section 3 contextualizes and reports on the challenges we faced and the lessons learned while empirically assessing software architecture technologies. Section 4 discusses the main limitations of the research, and Section 5 concludes the paper.

## 2 Motivation and Background

### 2.1 Study Motivation

In an industrial setting, when we compare the role of a software architect with that of a tester, our experience shows that people performing the former are senior software professionals, usually much older than people performing the latter. The main underlying principle is similar to the "Grand Old Duke of York" anecdote (Brown et al. 1998); in practice, it is necessary to have significant experience as an *implementationist* (i.e., programmer) to become a good *abstractionist* (i.e. architect). Confirming this observation is the fact that our students do not usually find employment as architects right after graduating. From this, we can assert that software architecture is mainly driven by experience.

Besides there are several challenges inherent in empirical research in software architecture, that we will describe in the next section, we note that there has been little interaction between the empirical software engineering community and the software architecture community. This situation has created a significant gap between these two communities. In particular, empiricists prefer studies with nice, closed, settings, and few variables, while architects do not see the applicability of such studies to large, long-lived software intensive systems. In fact, control and realism are both desirable characteristics of every empirical study but due to practical constraints, they may conflict with each other and trade-offs may be required (Murphy et al. 1999; Wohlin et al. 2000). In this context, the empirical community tends to prioritize control over realism, while it is the converse for the software architecture community. If there is a misalignment between *constructionists* and *empiricists* in the software engineering community (Erdogmus 2008), then it appears to be exacerbated in the software architecture field.

While there are a number of laws related to performance prediction (e.g., queuing networks) other quality attributes related to the process (analysis, design, and review), rather than to the product, lack the support of scientific laws, for example: customizability, scalability, and replaceability. This is exemplified in the comment "The life of a software architect is a long—and sometimes painful—succession of suboptimal decisions made partly in the dark" (Kruchten 2001). However, the experience gained over years of practice does help software architects to navigate in the dark. Books and collections of architectural design patterns are a further example of the evidential support available to architects; however, the patterns are mainly related to the product rather than to the process.

Without a supportive body of knowledge, architects base their decisions on commonsense and personal experiences only, rather than combining these aspects with sound evidence. On the one hand, empirical evidence is expensive and it is not normally ready to use, because practitioners need to understand the generalizability of the results and their context (Dybå et al. 2005). On the other hand, evidence can be a valuable source of knowledge. For example, in the study described in Section 3.1 regarding the size of teams for software architecture reviews, it was found that the quality of scenario profiles developed by a group did not increase linearly with the size of the group. Such information is expected to optimize the utilization of human resources allocated to software architecture reviews without having negative effect on the quality of the scenario profiles. In another study, described in Section 3.1, regarding the impact of documenting design rationale, with respect to the current practice of not documenting design rationale, it was found that in the presence of changes in requirements, individual and team decision-making had higher effectiveness when the design rationale documentation was available. Such information is expected to convince architects that they should document key decisions in order to avoid high maintenance costs, high degrees of design erosion, due to a lack of information and documentation of relevant architectural knowledge.

## 2.2 Software Architecture as a Discipline of Research and Practice

Researchers and practitioners have provided several definitions of software architecture and a list of definitions can also be found on Software Engineering Institute (SEI)'s website (SEI 2007). Since there is no standard, unanimously-accepted definition of software architecture, in this research, we use the most widely and commonly used definition of software architecture provided by Bass et al. in (2003): *"The software architecture of a program or computing system is the structure or structures of the system, which comprise software elements, the externally visible properties of those elements, and the relationships among them."* This definition is mainly concerned with the structural aspects of a system. Another commonly used software architecture definition that covers more than just the structural aspects, describes software architecture as a "set of significant decisions about the organization of a software system: selection of the structural elements and their interfaces by which a system is composed, behavior as specified in collaborations among those elements, composition of these structural and behavioral elements into larger subsystem, and the architectural style that guides this organization software architecture also involves usage; functionality; performance; resilience; reuse; comprehensibility; economic and technology constraints and trade-offs; and aesthetic concerns" (Kruchten 2003) based on (Shaw and Garlan 1996).

One of the main objectives of software architecture is to provide intellectual control over sophisticated systems of enormous complexity (Kruchten et al. 2006). And, over the last 15 years, software architecture has emerged as an important area of research and practice in the field of software engineering for managing the realm of large-scale, software-intensive systems development and maintenance (Clements et al. 2002a; Shaw and Clements 2006).

Why should we care about software architecture? Software architecture is developed during the early phases of the development process; it greatly constrains or facilitates the achievement of specific functional requirements, nonfunctional requirements, and business goals (Booch 2007b). In particular, a focus on software architecture supports risk mitigation, simplification, continuous evolution, reuse, product line engineering, refactoring, service-oriented engineering, acquisition, explicit expansion, systems of systems, and coordination (Booch 2007a).

Software architecture is an artifact; however, in our past studies we concentrated more on the supportive technologies developed to design, document, and evaluate software architecture.

Figure 2 describes software architecture design process as a whole; it is an iterative process with the following three phases:

1. **Understand the problem**: This phase consists of analyzing the problem and extracting the most critical needs from the big, ambiguous problem description. This phase is largely about *requirements analysis,* focusing on revealing those stakeholders' needs that are architecturally significant (Eeles 2005). This is done by determining the desired *quality attributes* of the system to be built, that, together with the business goals, drive the architectural decisions. The Quality Attribute Workshop (Barbacci et al. 2003) is an approach used for analyzing and eliciting architecturally significant requirements.

2. **Find a solution for the problem**: This phase consists of *decision-making* in order to fulfill the stakeholders' needs (as defined in the previous phase), by choosing the most appropriate architectural design option(s) from the available alternatives. In this phase, the properties of software components and their relationships are defined.

3. **Evaluate the solution**: Finally it is necessary to decide whether and to what degree the chosen alternative solves the problem. In the architecture context, this phase consists of *architectural evaluation*. Comprehensive descriptions related to this activity can be found in (Ali Babar and Kitchenham 2007a; Ali Babar et al. 2004; Dobrica and Niemelä 2002; Obbink et al. 2002).

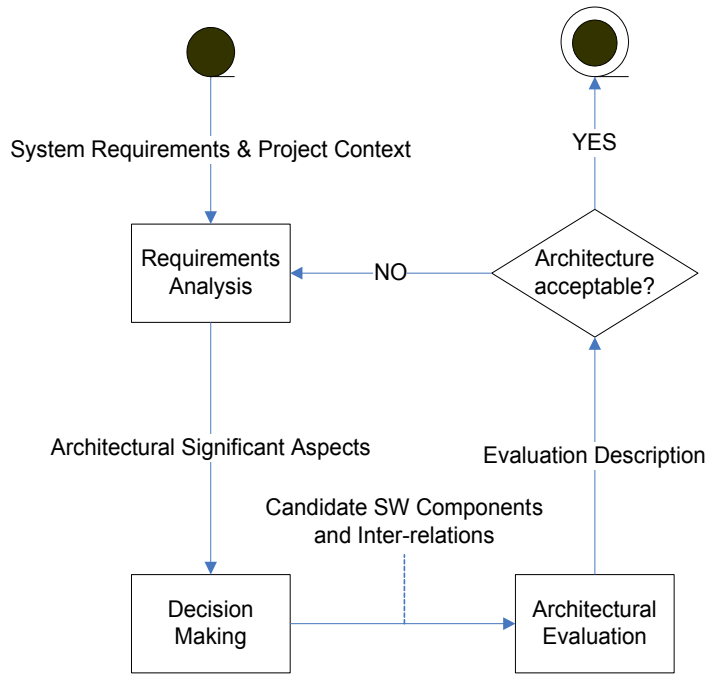

Fig. 2: The overall software architecture design phase.

Considering they were developed independently using different vocabularies existing software architecture design methods are very similar at the conceptual level (Hofmeister et al., 2007). Differences among software architecture design methods include the level of granularity of the decisions to be made, the concepts that must be taken into account, the emphasis on phases, the audience (large vs. small organization), and the application domain. A discussion regarding commonalities and variability of software architecture design methods can be found in (Hofmeister et al., 2007) and (Falessi et al., 2007), respectively.

**2.3 Related Studies**

The importance and current lack, of empirical assessment has been demonstrated in many software engineering areas like high performance computing (Shull et al. 2005), agile software development (Dybå and Dingsøyr 2008), regression testing (Engstrom et al. 2008), variability management (Chen et al. 2009), reverse engineering (Tonella et al. 2007), and information visualization (Ellis and Dix 2006). The Goal Question Metric (GQM) paradigm is a general approach for the "specification of a measurement system targeting a particular set of issues and a set of rules for the interpretation of the measurement data" (Basili et al. 1994). However, each software engineering area has its own problems for empirical assessment. We believe that each community should take the responsibility for building a body of knowledge in their respective areas of research and practice. Such an approach has provided excellent results in the area of software quality (Shull et al. 2006).

Ten years ago, Harrison Warren suggested that the lessons that empiricists learned "aren't the kinds of things you can write papers about (or at least papers that get published). In many cases they aren't significant enough, or general enough, or original enough, to make it through a rigorous refereeing process" (Harrison 1998). Meanwhile, the empirical software engineering paradigm has gained importance, as have the related lessons learned. The following paragraphs describe previous efforts that support the importance of reporting empirical experience, in the form of challenges and lessons learned, when building a body of knowledge related to the application of empiricism to specific software engineering areas.

Lung et al. in (2008) reported difficulties in validating the results of a previous study (Dehnadi and Bornat 2006) when replicating the study. In summary, they found different results even with minor changes in context. They claim that the main reason is that individual behaviour is difficult to replicate. A major cause could be due to differences among individual performances (Glass 2008).

Ji et al. in (2008) reported on their difficulties and lessons learned in conducting surveys in China on open source software and software outsourcing. In particular, they focused on addressing issues related to sampling, contacting respondents, data collection, and data validation.

Brereton et al. in (2007) reported on lessons learned in applying the systematic literature review method in software engineering. In particular, they report on lessons learned, from three studies, related to each of the ten stages of the systematic literature review method. They also report on certain inadequacies in the current publication system in its support of the application of the systematic literature review method. Their main findings were that infrastructure support provided by software engineering indexing databases is inadequate and the quality of abstracts is poor and not exhaustive. They report on their experiences regarding one empirical method and three study objects: service-based systems, the technology acceptance model, and guidelines for conducting systematic literature reviews. Staples and Niazi (2007) also report on systematic literature reviews when they discuss their experiences in following the guidelines for conducting systematic reviews as proposed in (Kitchenham 2004).

Desouza et al. in (2005) report on lessons learned in several software organizations when investigating post-mortem reviews as a viable method for capturing tacit insights from projects.

Shull et al. in (2005) provide guidelines and describe some of their experiences when designing controlled experiments for assessing high performance computing research. They also provide a web-based lab package with all the resources necessary for educators to do the same kind of study in their own courses.

Punter et al. in (2003) raise the awareness of online surveys and discuss the methods used to perform surveys in order to assess software engineering research. They also report on their experiences in performing online surveys in the form of lessons learned and guidelines.

Sjøberg et al. in (2003) report on the challenges and lessons learned in increasing the realism of controlled experiments related to object-oriented design alternatives. In particular, they explicitly highlight the importance of reporting in the literature challenges and lessons learned while empirically assessing software engineering methods. Hannay and Jorgensen recently improved these concepts in (2008).

Murphy et al. in (1999) report their experiences in empirically assessing aspect-oriented programming. They claim that their lessons learned are not only related to aspect-oriented programming but are also applicable to research in an early stage of development that attempts to assess new programming techniques.

With the aim of providing researchers with useful guidelines for carrying out experiments in software engineering, Basili et al. in (1986) present a framework for analyzing experimental studies. They highlight problem areas and lessons learned.

In conclusion, to the best of our knowledge, we provide the first attempt at reporting on challenges and lessons learned when applying empiricism to software architecture.

## 3. Experiences

### 3.1 Experimenting on software architecture technology

One of our main research goals is to advance the state of the art of the software architecture process (analysis, design, and review) by improving its supportive technologies (methods, techniques, and tools). To this end, we have conducted a series of empirical studies for assessing different software architecture related methods by following the principles of the evidence-based paradigm (Dybå et al. 2005). We emphasize that we have already reported the outcomes of our empirical studies extensively elsewhere (see references reported in Table 1); however, we have not always described our related experiences. We believe that sharing these insights is particularly valuable due to the increasing importance of software architecture and empiricism, and, above all, due to their potentially high interaction.

The empirical methods used in our research include controlled experiments (5), replicated experiments (3), expert opinion (1), literature reviews (2), and surveys (4), all involved as subjects both practitioners (360) and students (600); in the rest of this discussion when we place the numbers near empirical methods then they refer to the number of applications, when the number are near roles, then they refer to number of people performing that role. With our list of experiences, we aim to describe the sources of our experience. Easterbrook et al. in (2008) provide useful guidelines for selecting appropriate empirical methods for software engineering research. Table 1 presents some of the empirical studies that we have enacted on software architecture; each row represents a study, the different columns describe: the identifier of the study (S), the software architecture activity supported by the method being assessed (Activity), the main research question, the empirical strategy adopted, and the reference for further details. All studies were carried out by the authors and their research collaborators.

| S | Activity | Main Research Question | Empirical Strategy | Reference |
|---|---|---|---|---|
| 1 | Evaluation | Is there any difference in quality of scenario profiles created by different sizes of groups? | Experiment | (Ali Babar and Kitchenham 2007b) |
| 2 | Documentation | Does the documentation of design decision rationale improve decision making? | Experiment | (Falessi et al. 2006) |
| 3 | Documentation | Does the value of an information depend on its category and the activity it support? | Experiment | (Falessi et al., 2008a) |
| 4 | Documentation | Does the value of an information depend on its category and the activity it support? | Experiment replica | (Falessi et al. 2008b) |
| 5 | Design | Does a good code structure facilitate reengineering activity? | Pilot study + Experiment | (Cantone et al 2008b) |
| 6 | Evaluation | Is FOCASAM suitable to compare software architecture analysis methods? | Expert opinion | (Ali Babar and Kitchenham 2007a) |
| 7 | Design | Do software architecture design methods meet architects' needs? | Systematic Literature Review + Expert opinion | (Falessi et al. 2007a) |
| 8 | Evaluation | Does groupware-support-tool improve evaluation activity? | Experiment | (Ali Babar et al. 2008) |
| 9 | Evaluation | Does ALSAF support security sensitive analysis? | Pilot study + Quasi-expriment | (Ali Babar 2008) |
| 10 | Evaluation | Which factors do influence the architecture evaluation? | Focus group | (Ali Babar et al. 2007) |
| 11 | Documentation | How valuable is design rationale to practitioners? | Survey | (Tang et al. 2007) |

Table 1: Software architecture empirical studies.

The major contribution of Table 1 is to reveal empirical opportunities in software architecture by describing a large set of (addressed) research questions and the empirical strategy adopted. Further study details can be found in the references provided.

Table 2 reports on the relationship of our empirical studies with challenges and lessons learned. The columns refer to the studies enacted while the rows refer to the challenges or lessons learned. An *x* denotes that a given challenge or lesson learned was encountered in a particular study; an empty box denotes that a challenge or lesson learned was totally or mainly absent in a particular study.

|  |  | Empirical Studies |  |  |  |  |  |  |  |  |  |  |
|---|---|---|---|---|---|---|---|---|---|---|---|---|
|  |  | S1 | S2 | S3 | S4 | S5 | S6 | S7 | S8 | S9 | S10 | S11 |
| **Challenges (C)** | C1. Describing bounded rationality |  | x |  |  |  |  |  |  |  |  |  |
|  | C2. Describing other influencing decision |  | x |  |  |  |  |  |  |  |  |  |
|  | C3. Describing the desired ROI |  | x |  |  |  |  |  |  |  |  |  |
|  | C4. Describing social factors |  | x | x | x |  |  |  |  |  |  |  |
|  | C5. Describing the adopted SW arch. evaluation | x |  |  |  |  | x | x |  | x | x |  |
|  | C6. Evaluating the SW arch. without the system | x |  |  |  |  | x | x |  | x | x |  |
|  | C7. Cost of subjects |  | x | x | x | x |  |  |  |  |  |  |
|  | C8. Cost of review | x |  |  |  |  | x | x |  | x | x |  |
|  | C9. Cost of researcher |  |  |  |  |  |  |  | x |  |  |  |
|  | C10. Cost of training |  | x | x | x | x |  |  |  |  |  |  |
|  | C11. Complexity. | x | x | x | x | x |  |  | x |  |  |  |
|  | C12. Fuzzy boundaries |  | x | x | x |  |  |  |  |  |  | x |
|  | C13. Time bounded studies. | x | x | x | x | x |  |  | x |  |  |  |
| **Lessons Learned (LL)** | LL1. Contribution: methodology over results |  |  |  |  |  |  |  | x |  |  |  |
|  | LL2. Population: size over experience |  | x | x | x |  |  |  |  |  |  |  |
|  | LL3. Design: freedom over imposition |  |  |  |  | x |  |  |  |  |  |  |
|  | LL4. Execution: imposition over freedom |  |  |  |  | x |  |  |  |  |  |  |
|  | LL5. Objects: intended artificiality over aimed realism |  | x | x | x |  |  |  |  |  |  |  |
|  | LL6. Pilot studies for subjects and researchers |  | x |  |  |  |  |  |  |  |  |  |
|  | LL7. Pilot studies and replications |  |  |  | x |  |  |  |  |  |  |  |
|  | LL8. Interviews for triangulating results |  |  |  |  |  |  | x |  |  |  |  |
|  | LL9. Gathering qualitative data to explain quant. data | x |  |  |  |  | x |  |  |  |  |  |
|  | LL10. Attracting practitioners as participants |  |  |  |  |  |  |  |  | x | x | x |

Table 2: Relations of enacted empirical studies with challenges or lessons learned.

Like Brereton et al. (2007), in order to contextualize the challenges and lessons learned, we describe our empirical studies with the structured abstract headings: context, objectives, method, results and conclusions. Due to space constraints, we describe only Study 1 and Study 2 because they are most related to the challenges and lessons learned reported below (see Table 2).

### S1: The impact of group size on evaluation

*Context and study motivation*: Architecture evaluation involves a number of stakeholders working together in groups. In practice, group size can vary from two to twenty stakeholders. Currently there is no empirical evidence concerning the impact of group size on group performance. Hence, there is a need to explore the impact of group size on group performance for software architecture evaluation.

Falessi, D., Ali Babar, M., Cantone, G., Kruchten, P., Empirically Assessing Software Architecture Research: Challenges and Lessons Learned, Empirical Software Engineering Journal, 15(3): pp. 250-276, 2010.

*Objectives*: The main objective of this study was to gain some understanding of the impact of group size on the outcome of a software architecture evaluation exercise. Initially, we decided to explore the impact of group size on the scenario development activity. This study was intended to find answers to the following research questions: (1) Is there any difference in quality of scenario profiles created by different sizes of groups? and (2) How does the size of a group affect the participants' satisfaction with the process and outcomes, and their sense of personal contribution to the outcome?

*Method*: This experiment compared the performance of groups of varying sizes. The experiment which used a randomized design, used the same experimental materials for all treatments. Subjects were randomly assigned to groups of three different sizes (3, 5, and 7). The independent variable in this study is the size of a group (number of members) and the dependent variable is the quality of scenario profiles developed by each sized group. The questionnaire gathered participants' demographic data and information on their satisfaction with the meeting process, quality of discussion, and solution, and commitment to and confidence in the solution.

*Results and conclusions*: Analysis of the quantitative data revealed that the quality of scenario profiles for groups of 5 was significantly greater than that for groups of 3, but there was no difference between groups of 3 and 7. However, participants in groups of 3 had a significantly better opinion of the group activity outcome and their personal interaction with their group than participants in either groups of 5 or 7. From these findings we can conclude that the quality of the output from a group does not increase linearly with group size. However, individual participants prefer small groups. These findings were consistent with the results of studies on optimum team size for software inspections, where researchers agree that the benefits of an additional inspector diminish with growing team size (Biffl and Gutjahr 2001). These findings provide the first empirical evidence that supports having relatively smaller teams for architecture evaluation. The findings from this experiment also enabled us to propose a new format for architecture evaluation in geographically software development distributed teams by leveraging the empirical findings of our previous studies; these studies revealed that geographically dispersed teams can be more effective than collocated teams, although individual participants preferred face-to-face meetings.

## *S2: The Impact of Design Decision Rationale Documentation*

*Context and study motivation*: Despite the fact that individual and team decision-making have a crucial influence on the level of success of any software project, few empirical studies have evaluated the utility of design decision rationale documentation. Although several studies consider approaches and techniques for documenting design decision rationale and have argued about the benefits only one study has focused on performance and has evaluated this in a controlled environment.

*Objectives*: The aim of this research is to experimentally evaluate the Decision Goals and Alternatives (DGA) method for documenting design rationale when compared with the current practice of not documenting design rationale at all. If we present our objective more formally and according to the GQM template (Basili et al. 1994), then the goal of the study presented here is to empirically analyze the DGA technique (Falessi and Becker 2006), with respect to the effectiveness and efficiency of individual-decision-making and team-decision-making, in case of changes in requirements, from the point of view of the researcher, in the context of post-graduate Master students of software engineering.

*Method*: We conducted a controlled experiment at the University of Rome "Tor Vergata", with fifty Masters students performing in the role of experiment subjects. Design decisions regarding an ambient intelligence project prototype developed at Fraunhofer IESE (ISESE 2008) constituted the experimental objects. The context of the study is off-line (an academic environment) rather than on-line, based on students rather than professionals, using domain-specific and goal-specific real objects rather than generic or toy-like objects (Wohlin et al. 2000).

*Results and conclusions*: The main results of the experiment are derived from objective data and show that, with changes in requirements, individual and team decision-making are as follows: (1) Whatever the kind of design decision effectiveness improves when DGA documentation is available. (2) the DGA documentation did not appear to affect efficiency. Supplementary results based on subjective data, allowed us to confirm, with triangulation, the main results regarding the utility of DGA.

## 3.2 Challenges

This subsection reports on, in separate paragraphs, the challenges encountered. We note that the challenges described below may be relevant and applicable to several software engineering fields; however, we believe that they are exacerbated in the software architecture field.

In general, the empirical paradigm assesses a method by measuring its performance, when it is used by people. Such an assessment can focus on the product (e.g., number of defects), the process (e.g., required effort), and the resource (e.g., subjects' age) (Wohlin et al. 2000). Therefore, if we are interested in comparing two technologies that each supports the software architecture process, it is relevant to compare the quality of the derived architectures. Hence, even when the architecture evaluation activity is not the activity being assessed, such an activity must be enacted to support the empirical investigation. Consequently, despite the fact that most of the challenges mentioned below are related to the software architecture evaluation activity, we believe that they are also relevant to other software architecture process activities, for instance design and documentation. Our description of the challenges is divided into three subsections: measurement control, investigation cost, and object representativeness.

### *3.2.1 Measurement Control: Objectively Measuring Software Architecture Goodness*

The GQM approach (Basili et al. 1994) provides a generic and systematic way of defining a suitable set of metrics for a given context. However, defining the level of goodness of software architecture is a complicated matter. According to Bass et al. (2003), "analyzing an architecture without knowing the exact criteria for goodness is like beginning a trip without a destination in mind." Booch states that "one architectural style might be deemed better than another for that domain because it better resolves those forces. In that sense, there is a goodness of fit—not necessarily a perfect fit, but good enough" (Booch 2006b).

In the following section, we describe the challenges encountered in measuring the goodness of a software architecture, when such a measurement is required as a criterion for assessing a given technology designed to support the software architecture process. Difficulties in describing the factors that influence the goodness of a given software architecture constitute a barrier for a researcher trying to measure and/or control related empirical variables at a constant level (e.g., according to Tom De Marco, "you cannot control what you cannot measure," (De Marco 1986). This means that if there is something we are not able to describe/identify in advance, then we cannot be sure that the results of the empirical study conducted depend on the defined treatment(s) and not on something else.

**C1. Describing bounded rationality.** The level of goodness depends heavily on the amount of knowledge available at evaluation time (Simon 1996). Software architecture is an artifact that is usually delivered at a very early stage in the software development lifecycle. This means that software architecture decisions are often made with unstable and quite vague system requirements. Hence, software architecture goodness depends on the existing level of risk for incomplete knowledge, which is difficult to describe and hence analyze as an impact factor. In other words, some supportive technologies, like the rationale documentation assessed in S2, may support to varying extents,, the architecture process depending on the level of knowledge of the architect (which is hard to measure).

**C2. Describing other influencing decisions.** Design decisions are made based on the characteristics of the relationships that they have with other decisions, which are outside of the architect's researching range (see *pericrises* by Kruchten in (2004)). Since impacts amongst decisions are hard to control, then the goodness of a decision is difficult to measure. In order to cope with this challenge, in S2 we describe decisions relationships using the framework proposed by Tyree and Akerman in (2005).

**C3. Describing the desired Return on Investment.**, For the development of any system, the optimal set of decisions is usually the one that maximizes the Return on Investment (ROI). |With such a view, for instance, an actual architecture might be considered more valuable than a better potential one achievable by applying some modifications to the existing architecture; in fact, the potential architecture would introduce additional risk, may delay project delivery, and might result in financial loss. In practice therefore, ROI is an important factor in defining the goodness of a software architecture. However, desired ROI changes over time and it is difficult to precisely describe. In S2, we carefully described timing we wanted when maximizing the return in making the decision.

**C4. Describing social factors.** Social issues such as business strategy, national culture, corporate policy, development team size, degree of geographic distribution, etc., all can significantly influence the design decision making process. Therefore, although social factors may influence the goodness of an architecture they are difficult to report due to several factors like nondisclosure agreements or implicit assumptions. We experienced this challenge during technology transfer.

**C5. Describing the adopted software architecture evaluation.** It can be assumed that different software architecture evaluation approaches may lead to different results unless there is strong evidence otherwise. Ali Babar et al. (2004) have proposed a set of attributes for characterizing different software architecture evaluation methods. The set of attributes represents a basic frame of reference for comparing different architecture evaluation methods. Additionally, in evaluating software architecture, we assume that different types of input may lead to different results. The nature and number of inputs varies depending upon the kind of architecture evaluation method used. Several researchers and practitioners have proposed different sets of inputs as reported in (Clements et al. 2002b) and (Obbink et al. 2002). In conclusion, the evaluation step is difficult to describe comprehensively (i.e. to be replicable); this provides a further barrier in the application of rigorous empirical approaches when evaluating software architecture technologies.

**C6. Evaluating the software architecture without analyzing the resulting system.** Large complex software systems are prone to be late to market, they often have quality problems, and provide fewer functionalities than expected (Jones 1994). It is important to uncover any software problems or risks as early as possible. Reviewing the software architecture is a valid means to check system conformance and to reveal any missed objectives early in the development lifecycle (Maranzano et al. 2005) because: (1) software architecture is developed during the early phases of the development process, and (2) it constrains or facilitates the achievement of specific functional and nonfunctional requirements, and business goals. Hence, a software architecture review can be an effective means to predict the *ilities* of the resulting system (Obbink et al. 2002) (Kazman et al. 2004) like performance (Liu et al. 2005) and modifiability (Bengtsson et al. 2004). However, since such predictions cannot be perfectly accurate, the resulting system may not be able to achive the desired and predicted property levels. This occurs because architectural decisions constrain other decisions (e.g., detailed design, implementation), which also impact system functionalities. Architectural decisions interact with each other (Kruchten 2004) (Eguiluz and Barbacci 2003); "The problem is that all the different aspects interrelate (just like they do in hardware engineering). It would be good if high-level designers could ignore the details of module algorithm design. Likewise, it would be nice if programmers did not have to worry about high-level design issues when designing the internal algorithms of a module. Unfortunately, the aspects of one design layer intrude into the others." (Reeves 1992).

*3.2.2 Investigation Cost*

From an industrial point of view, an empirical study is considered an investment that is made in order to produce a return (Prechelt 2007). From a research institute/academic point of view, the limitation is the resources available for a study. Therefore, in every case, the cost required to run a study is an important factor in its selection and design. In the following, we describe some aspects that make the empirical assessment of software architecture a quite expensive undertaking.

**C7. Subjects.** In general, software architecture decision making requires a high level of experience. This is due to some facts already mentioned: architecture design provides the blueprint for the whole system, it hugely constrains or facilitates the achievement of specific functional and nonfunctional requirements, and business goals (Booch 2007b). Therefore, architects needs to consider several trade-offs, technological as well as organizational and social. In this context, using empirical subjects with little experience (e.g., students) may not be considered to be representative of the state of the practice in software architecture. But let us note that this is not a specific limitation of software architecture studies. For instance, studies on pair programming show different results from experiments using professionals (Arisholm et al. 2007) and those using students (Williams and Upchurch 2001). Nevertheless, many empirical software engineering academic studies recruit students and academics as experimental subjects to perform the role of software architect as in S2, S3, S4, and S5; it is still unclear whether it is reasonable, and to what extent academics can be considered able to sufficiently function in the role of software architect. However, experienced subjects are an expensive resource, whose cost is a significant barrier to carrying out empirical studies.

**C8. Reviews.** Reviewing software architecture is quite a complex task which is why it requires a lot of experience in the domain. Consequently, an architecture review is an expensive task. According to Bass et al. (2003), a professional architecture review costs around 50 staff days. Of course, such a cost is a large barrier to carrying out a well designed rigorous empirical study of a particular technology to support the software architecture review process.

**C9. Researchers.** The design, execution, and report of high-quality empirical studies requires much effort and many resources. We have observed that this aspect of empirical research into software architecture is usually underestimated by most researchers. Failure to correctly estimate the effort and resources required by a research team may result in a weak study and inconclusive or unreliable findings. Our experience is that the preparation of the design and other materials for a controlled experiment can take up to 3000 hours, depending upon the nature of the study. For example, the study reported in S8 took around 2800 hours of work, just for planning and material preparation. Planning a focus group and inviting participants can take a painstakingly long time, for which a researcher needs to be prepared. In our experience, the effort required from researchers for effectively preparing the materials and planning the execution of an empirical study is a commonly underestimated factor; therefore, the availability of the researchers' time becomes a problem A further challenge in designing and conducting high quality empirical studies regards the training and expertise of researchers, both in empiricism and the required software architecture topics,.

**C10. Training**. The participants of an empirical study on the use of a particular technique are expected to have a good knowledge of the concepts underpinning that technique (e.g., pattern-based evaluation or perspective-based readings in inspection). Software architecture concepts and principles cannot be taught in short training sessions to even practitioners with substantial experience in software development, let alone to university students. Hence, it is a challenge for an empiricist to determine the amount and duration of training required for the participants of an empirical study. This challenge puts pressure on the resources required for carrying out an empirical study — the more time required for training the less likely the participants will be available for the study.

*3.2.3 Object Representativeness*

The realism and representativeness of the objects adopted in software engineering studies have been promoted as an important means of increasing generalizability and industrial relevance (Houdek 2003; Laitenberger and Rombach 2003; Sjøberg et al. 2003). The idea supporting this argument is that empirical results are generalizable when the context studied is similar to industrial situations. However, there appears to be a consensus among several researchers that "deliberately introduced artificial design elements may increase knowledge gain and enhance both generalizability and relevance" (Hannay and Jørgensen 2008). The following paragraphs describe the problems we have faced in the construction of artificial empirical objects.

**C11. Complexity.** One of the main intents of software architecture is to provide "intellectual control over a sophisticated system's enormous complexity" (Kruchten et al. 2006). Hence, software architecture, as a discipline, is really useful only for large software systems whose complexity would not be otherwise manageable. The use of software architecture artifacts for small or simple systems, like the empirical objects that are frequently adopted in academic studies with students, is not representative of the state of the practice. Such studies neglect the phenomena characterizing complex systems. In other words, the results concerning the use of software architecture artifacts for *toy* systems do not scale up because the design of large complex system involves issues that are rarely experienced in the design of *toy* systems. This constitutes a barrier to the construction of valid artificial empirical objects as the results from empirical studies using *toy* systems have severe limitations.

**C12. Fuzzy boundaries.** There is no clear agreement on a definition of a software architecture (Smolander 2002) (SEI 2007). Software architecture encompasses the set of decisions that have an impact on the system behavior as a whole (and not just parts of it). Hence, an element is architecturally relevant based on the locality of its impact rather than on where or when it was developed (Eden and Kazman 2003). The difficulty in specifying the boundaries between software architecture and the rest of the design is a barrier to the selection of valid empirical objects to study. In S2 the decisions adopted were driven by major business goals and non-functional requirements.

**C13. Time bounded studies.** There is usually a limitation on the time available for conducting an empirical study (e.g., a controlled experiment or interview). It is difficult to convince practitioners to allocate enough time to carry out a study on a realistic problem. Academic studies are usually done in scheduled laboratory sessions that usually last between 1 and 2 hours. Hence, a researcher needs to provide or find a study object, like that in S2, which is not only small enough to be studied in the given time slot, but also real enough to make the results reliable and generalizable.

### 3.3 Lessons Learned

During the past years, while facing all the abovementioned challenges, we have learned a set of lessons. The aim of this subsection is to report on these lessons in order to provide valuable input for future empirical assessments.

**LL1. Contribution: methodology over results.** All the challenges presented in Section 3.2 can threaten the validity of the results of empirical studies of software architecture. However, the contribution of an empirical study is not only in its results, which aim to be generalizable, but also in the empirical approach used, which also aims at being replicable. Empirical approaches are becoming increasingly important when assessing the outcomes of software architecture research. Hence, empirical approaches need to be carefully designed during the study preparation so that they deal appropriately with the challenges they will encounter, and carefully reported afterwards to support replications. In some of our controlled experiments, where the main contribution was the results (supposed to be generalizable), faced difficulties in their reporting as reviewers were critical of the value of the results in terms of generalizability. However, one of our pilot studies, where the main contribution was the assessment of the suitability of the empirical methodology being used, was published as a journal paper like S8. From these experiences, we found that the solid and appropriate use of an empirical methodology is appreciated. While the results are of course valuable, we claim that the methodology is often underrated by the audience, especially practitioners. As a matter of fact, this particular challenge poses a particularly high threat to validity, and that in turn should shift the focus of the audience from the results to the methodology, when assessing software architecture research.

**LL2. Population: size over experience.** The issue of using students as subjects in empirical studies have been described in (Carver et al. 2003). Generally, it is obvious that people with the same level of expertise tend to act similarly; therefore, using students may inhibit generalizability (Potts 1993) (Glass 1994). Sjøberg et al. in (2003) provide guidelines for increasing the realism in controlled experiments. However, researchers should also be aware of the enormous cost associated with increasing the realism. Sometimes the level of realism required can be achieved with well-trained student participants. While considering different aspects of transferring the results from some of our experiments to practitioners, we have identified four main issues when using students as subjects:

1) *Evidence*: There are some indicators where the differences in performance between students and practitioners may not be relevant; examples are (Svahnberg et al. 2008) and (Host et al. 2000) in the context of requirements selection and assessment of lead-time impact. However, the results achieved with student participants are usually considered not generalizable by practitioners to their conditions unless there is solid supporting evidence otherwise.

2) *Experience*: Most of computer science and software engineering courses include practical exercises or projects delivered against preset deadlines. In addition, many students are expected to gain industrial experience during their third or fourth year of studies. We have also observed that a large number of students start working part-time as programmers or in technical support roles during their final years of undergraduate studies. Sjøberg et al. in (2001), have also suggested that graduate computer science students should be considered to be semi-professionals and hence are not so far from practitioners. However, we admit that on the other hand, there are too many graduate students, doing a Masters or Ph.D., that have never ever set foot anywhere near the real world. The danger is that they consider themselves as experts, and look upon seasoned practitioners with contempt.

3) ***Heterogeneity***: individuals' performance may vary hugely (Glass 2008), and professionals tend to vary more than students. Therefore, "the variations among students and variations among professionals may be so large that whether the person is a student or a professional, may just be one of many characteristics of a software engineer" (Sjøberg et al. 2002).

4) ***Sample size***: since the cost of subjects increases according to both their number and their experience, using inexperienced subjects allows the use of a large population. The benefit of using a large sample is twofold, it supports:

- statistical analysis: a large sample size increases the power of a significance test and also helps fulfill some of the requirements of using parametric tests.

- generalizability of results by inhibiting the effects of individual peculiarities: as we have already noted, the performance of humans varies a lot; therefore, the larger the sample size, the greater the results' generalizability.

In conclusion, while the degree of subjects' experience is of course valuable, we believe that the value of the population size is usually underrated by many, especially practitioners. Generalizability of results can be increased both with a larger sample size and with more experienced participants. However, due to the existence of constraints, the ideal way is a trade-off between these two factors.

In the following, we report on a strategy, as applied in S2, S3 and S4, for maximizing students' experience and hence increasing the generalizability. In fact, in S2, S3, and S4, we did not have the opportunity of using professionals so we had to use Masters students as subjects. However, we noticed that, on average, the students had a specific IT specialty, due to personal interests, academic vitae, and/or some industrial experiences. To emulate the context of real world decision-making, we tried to maximize their experience by designing the experiments in following way:

(1) We designed five different roles for the participants, one for each of the following areas: hardware; communication; software architecture and services discovery; interface; and data storage,

(2) well in advance of the last training session, subjects expressed their preference for each role, according to their previous experience and level of confidence with the role's responsibilities, and

(3) we assigned subjects to roles by maximizing the total according to the expressed preferences.

In this way, the subjects performed tasks in which they were experienced, or at least purported to be. We believe that such an approach significantly helped us to achieve realism.

**LL3. Design: freedom over imposition.** S5 regards a controlled experiment with the aim of *analyze* the Model View Controller (MVC) (Booch 2006a) design pattern, *for the purpose of* evaluating the impact, *with respect to* the effort required to develop and maintain a medium size application, toward a web-services system architecture, *from the point of view of* the researcher*, in the context* of fifty graduated Masters students playing the role of subjects. Our results showed that, on average, the people assigned to adopt an implicit architecture, rather than MVC, performed better (hence, in some sense, it applied a *better* structure) with respect to both the development and maintenance phases.

The main lesson learned was that design decisions (in our case a specific code structure) should not be imposed *a priori*. To the contrary, developers should have an awareness (Vokac et al. 2004) of the available solutions rather than impositions. The decision making process for selecting among design decisions should take into consideration: (1) the experience of the developers, (2) the characteristics of the specific business goals, and (3) current context peculiarities like application complexity. For instance, in order to emulate real software, in (Cantone et al. 2008), we adopted a medium size application as the empirical object to develop, which resulted in 20 KSLOC. While this is not really minor in size, it is not the only factor that needs to be taken into consideration while increasing realism: in fact, because of the low complexity of the application as a whole, we would classify it as *toy* software. As already reported by (Vokac et al. 2004), design patterns do not pay off in the case of *toy* applications. We conclude that a higher realism would be achieved by adopting an application of similar size but with real business goals and end users.

**LL4. Execution: imposition over freedom.** In a controlled experiment, different groups of subjects apply specific treatments. Afterwards, the treatments are assessed by comparing the performances of the different groups in terms of dependent variables. Since, there is always the possibility that subjects do not apply the assigned treatments, researchers are generally encouraged to ensure the proper application of the treatments. However, when assessing software architecture design, the treatments may be code structures to which checking conformance is not trivial. In S5 we had had two groups, one assigned to apply the MVC pattern, the other the implicit architecture (i.e., not care about code structure). In that study, we were not able to check the application of the treatment. Therefore, when we examined the empirical objects, we were not sure to what extent the MVC group really applied the MVC pattern. On the other hand, some subjects assigned to not care about code structure might have applied MVC to some extent. Hence, since we were not sure about that data partitioning, as derived from the nominal partition of participants in MVC-architecture subjects and the implicit architecture subjects, respectively, there was a risk that some data should move from the MVC to the implicit architecture group, and vice versa.

**LL5. Objects: intended artificiality over aimed realism.** Reproducing architectural objects in a synthetic setting is sometimes unfeasible in the software architecture context (see challenges C1, 2, 3, 4, 12, 13). Therefore, in the case of a synthetic setting (e.g., controlled experiment), it may be better to intentionally introduce some artificial elements rather than ineffectively trying to duplicate reality (Hannay and Jørgensen 2008). For example, in S2, S3, and S4, the projects were described but not implemented. That kind of project description produced a system that was sufficiently detailed and complex to use as the locus of the objects in the experiment. In fact, applying experimental tasks was non-trivial because subjects had to re-make decisions based on several opposite and inter-related objectives that characterized those decisions. This is how it is in the real world. As a further example, again in S2, S3, and S4, the key idea was to use single decisions as the experimental objects. This is not contradictory to the current trend to consider software architecture as a set of design decisions (Kruchten 2003) (Jansen and Bosch 2005). Hence, our preference was for analyzing the performance of software engineering methods by using one decision at a time rather than the whole set of decisions together. We found that breaking down the decision process was a positive action that provided more control and replicability.

**LL6. Pilot studies for subjects and researchers**. Software architecture is abstract in nature (see C13 in Section 3.2). That is why developing effective tasks and instrumentation is particularly difficult. Therefore, researchers need to have confidence specifically with instrumentation and tasks. Running a pilot study provides an effective way to let subjects get experience with instrumentation and tasks (independently from the knowledge taught during the training sessions) and for researchers to identify any problems in study design and experimental material and tasks. For instance, in the pilot study that we ran before the experiment described in S2, we noticed that the subjects who were charged with recording the amount of time spent performing a given task were inclined to greatly round off the data. To gather fine-grained data, we asked the subjects to write the actual time just before starting and after completing a task. Afterwards, we easily computed the required time by subtracting the two data sets. This change provided us with more accurate and fine-grained data.

**LL7. Pilot studies and replications.** Replications usually require much detailed information; replication packages provide a valid means to enhance communication between researchers (Vegas et al. 2006). However, since it is generally difficult to predict which information needs to be included in the package, researchers cannot be sure which tacit knowledge influences the replication results (Shull et al. 2002). In S4 we experienced many difficulties in replicating the previous study (S3); this occurred because architecture is abstract in nature (see the abovementioned challenges C1, 2, 3, 4, 12, 13). Hence, from S4 we learned that running a (even very short) pilot study pays off, even in case of exact replication because the latter still contains some novelty, as for instance the subjects' experience and the translated documentation. The role of novelty in replicated experiments is described in (Brooks et al. 2008; Kitchenham 2008). In particular, despite the fact that in S4 we enacted an exact replication, we rechecked the conformance of the instrumentation by letting the experimental subjects to try it and provide us with feedback.

**LL8. Interviews for triangulating results.** The software engineering community has developed a plethora of approaches, each with their own ontology. The software architecture artifact concerns many different methods. While comparing different design methods in S7, learned that there can be a wide variety of terminologies. The same experience was reported to us from people involved in the definition of the standards for "systems and software engineering architectural description" (ISO/IEC 42010 2008) a joint IEEE, ISO revision of the recommended practice for architectural description of software intensive systems. In S7, we confirmed our literature review results by directing the subjects interviewed. We discovered that while at times similar concepts are referred to by different terms (e.g., *use case* and *user story*), on other occasions the concepts do not overlap (e.g., for specific software architecture views). We learned from S7 that interviewing the authors to check on proper terminology understanding enhances the internal validity of an empirical study.

**LL9. Gathering qualitative data to explain quantitative data**. We have already discussed the challenges involved in getting sufficient number of participants with a desirable level of experience for empirical studies when assessing software architecture research. This is one reason for obtaining the maximum information from an empirical study. We believe that it is very good practice to obtain self-reported qualitative as well as quantitative data. The self-reported qualitative and quantitative data can provide additional explanatory information to assist with interpretation of the results achieved from the analysis of experimental data. Analysis of the self-reported data provides useful insights into initial observational studies. To this end, in S1 and S6 we learnt that using post-study questionnaires is quite effective in providing additional information about the participants' experiences, opinions, and attitudes towards a particular treatment or control.

**LL10. Attracting practitioners as participants**. While we cannot overemphasize the importance of empirical studies involving practitioners as participants, we have already discussed the challenges involved in getting practitioners involved in empirical studies of software architecture. We learned from S9 that practitioners can be enticed to participate if it has some value to them in terms of training and learning on a topic that interest them. Additionally, the response rate of a survey can be improved if potential respondents are confident that the accumulated results of the study could enable them to benchmark their practices with their counterparts in other companies. We have used these strategies in all of our survey studies into different aspects of software architectures as reported in S11 and in focus group like S10. All of these survey studies have achieved a reasonably good response rate. Other researchers have tried to recruit practitioners for their control experiments by paying according to the cost of their time (e.g. (Dzidek et al. 2008)). However, such studies are extremely expensive to carry out and even then the availability of practitioners is not guaranteed. We have found that offering training as an incentive for participating in research is quite an attractive factor not only for the practitioners concerned, but also for their organizations.

## 4. Limitations

The major limitation of this paper is that the challenges and lessons learned and reported here come only from our collective experience. This means that they may have been exacerbated or inhibited by:

- Our personal attributes, such as our level of experience, of expertise, and of ability in designing and conducting empirical studies.
- The specific Software Architecture technologies that we chose to assess.
- The specific empirical methodologies that we used in our research.

Consequently, there may be different challenges or more effective solutions than the ones we have reported in this paper. However, the challenges we have provided and the lessons learned are real, and they come from a large spectrum of empirical methodologies, software architecture technologies, and collaborations with several researchers around the world.

A common limitation of empirical research is that the technology assessed is developed and assessed by the same group of researchers. In such a context, it can be expected that researchers may consciously and/or unconsciously influence the results. In this research, we believe that the content is unlikely to be biased because there are no benefits in reporting challenges or lessons learned that we have not faced, or have not observed in the empirical studies on software architecture conducted by us, our colleagues and our peers.

## 5. Conclusion

The novelty of this paper lies in its characterization of the empirical paradigm with respect to its applicability to software architecture. In fact, over the past years, software architecture researchers have been very active in developing new methods, techniques, and tools in order to support the architecture evaluation process; however, a majority of these technologies await rigorous empirical assessment. We believe that without systematically accumulating and widely disseminating evidence about the efficacy of different methods, techniques, and tools it would be naive to expect successful technology transfer and improvement. Anecdotal evidence alone, irrespective of the credibility of the source, may not be enough to convince organizations to include a technology in their portfolio and train employees to use it. Since empiricism provides scientifically valid approaches to systematically gather and use evidence, the aim of this paper is to promote and facilitate the application of empiricism to software architecture research. To this end, we have described 13 challenges and 10 lessons learned that we have actually experienced in assessing software architecture research through applying controlled and replicated experiments, expert opinion, systematic literature review, observation study, and surveys.

We claim that the challenges described above should not act as inhibitor; software architecture researchers must follow a two-pronged strategy: develop new techniques to improve on current practices, and perform systematic, rigorous assessments of existing and new techniques by following the empirical paradigm. As a matter of fact, it is by focusing on the existence of these challenges that researchers can more accurately plan their assessment studies without overlooking significant empirical aspects. As some software engineering areas are more mature than other areas from an empirical perspective (e.g., software testing vs. software architecture), we believe that the validity-threshold for publication review should be defined by taking into account the actual maturity of a given field. Such an approach would allow a field to incrementally mature from an empirical perspective. In other words, it would be naive i) to compare the validity of an empirical study on software architecture to one on software testing and ii) to

Falessi, D., Ali Babar, M., Cantone, G., Kruchten, P., Empirically Assessing Software Architecture Research: Challenges and Lessons Learned, Empirical Software Engineering Journal, 15(3): pp. 250-276, 2010.expect that the empirical maturity of a given software engineering field would evolve other than step by step.

In conclusion, a greater synergy between the empirical and software architecture communities, as suggested and fostered by this paper, would support:

1) the emergence of a body of knowledge consisting of more widely-accepted and well-formed theories on software architecture,

2) the empirical maturation of the software architecture field by allowing software architecture researchers to share their empirical experiences, for example, in terms of lessons learned; this would in turn promote the point 1 above.

## Acknowledgement

The authors wish to thank the colleagues involved in the empirical studies from which the reported challenges and lessons were drawn, Professor June Verner for helping in the preparation of the proof version, and the anonymous reviewers for their very insightful comments and helpful suggestions.

Falessi, D., Ali Babar, M., Cantone, G., Kruchten, P., Empirically Assessing Software Architecture Research: Challenges and Lessons Learned, Empirical Software Engineering Journal, 15(3): pp. 250-276, 2010.

# References

Ali Babar M., Zhu L., Jeffery R. (2004) A framework for classifying and comparing software architecture evaluation methods. Proceedings of the Australian Software Engineering Conference. .

Ali Babar M., Bass L., Gorton I. (2007) Factors influencing industrial practices of software architecture evaluation: An empirical investigation. Quality of Software Architectures (QoSA). Massachusetts,USA.

Ali Babar M., Kitchenham B. (2007a) Assessment of a framework for comparing software architecture analysis methods. 11th International Conference on Evaluation and Assessment in Software Engineering (EASE)

Ali Babar M., Kitchenham B. (2007b) The impact of group size on software architecture evaluation: A controlled experiment. Proceedings of the First International Symposium on Empirical Software Engineering and Measurement. IEEE Computer Society.

Ali Babar M. (2008) Assessment of a framework for designing and evaluating security sensitive architecture. 12th International Conference on Evaluation and Assessment in Software Engineering (EASE08). Bari, Italy.

Ali Babar M., Kitchenham B., Jeffery R. (2008) Comparing distributed and face-to-face meetings for software architecture evaluation: A controlled experiment. Empirical Software Engineering; An International Journal 13 (1), 39-62.

Arisholm E., Gallis H., Dybå T., Sjoberg D. (2007) Evaluating pair programming with respect to system complexity and programmer expertise. IEEE Transaction on Software Engineering 33 (2), 65-86.

Barbacci M. R., Ellison R., Lattanze A. J., Stafford J. A., Weinstock C. B., Wood W. G. (2003) Quality attribute workshops (qaws), third edition.
http://www.sei.cmu.edu/publications/documents/03.reports/03tr016.html

Basili V., Selby R., Hutchens D. (1986) Experimentation in software engineering. IEEE Transactions on Software Engineering 12 (7), 11.

Basili V., Caldiera G., Rombach D. (1994) Goal/question/metric paradigm. Encyclopedia of Software Engineering 1 (John Wiley & Sons), 528-532.

Basili V. R. (1996) The role of experimentation in software engineering: Past, current, and future. Proceedings of the 18th International Conference on Software Engineering. Berlin, Germany. IEEE Computer Society.

Bass L., Clements P., Kazman R. (2003) Software architecture in practice. 2nd Addison-Wesley. Reading, MA.

Bengtsson P., Lassing N., Bosch J., Vliet H. v. (2004) Architecture-level modifiability analysis (alma). Journal of Systems & Software 69 (1-2), 129-147.

Biffl S., Gutjahr W. (2001) Influence of team size and defect detection technique on inspection effectiveness. Proceedings of the 7th International Symposium on Software Metrics. IEEE Computer Society.

Biffl S., Aurum A., Bohem B., Erdogmus H., Grünbacher P. (2005) Value-based software engineering. Springer.

Boehm B. W. (1981) Software engineering economics Prentice Hall PTR Advances in computing science and technology. Prentice-Hall. Englewood Cliffs, N.J.

Booch G. (2006a) The accidental architecture. IEEE Software 23 (3), 9-11.

Booch G. (2006b) Goodness of fit. IEEE Software 23 (6), 14-15.

Booch G. (2007a) The economics of architecture-first. IEEE Software 24 (5), 18-20.

Booch G. (2007b) The irrelevance of architecture. IEEE Software 24 (3), 10-11.

Brereton P., Kitchenham B. A., Budgen D., Turner M., Khalil M. (2007) Lessons from applying the systematic literature review process within the software engineering domain. Journal of Systems and Software 80 (4), 571-583.

Brooks A., Roper M., Wood M., Daly J., Miller J. (2008) Replication's role in software engineering. in. Guide to advanced empirical software engineering Springer.

Brown W. J., Malveau R. C., McCormick H. W., Mowbray T. J. (1998) Antipatterns: Refactoring software, architectures, and projects in crisis. Wiley.

Falessi, D., Ali Babar, M., Cantone, G., Kruchten, P., Empirically Assessing Software Architecture Research: Challenges and Lessons Learned, Empirical Software Engineering Journal, 15(3): pp. 250-276, 2010.

Cantone G., D'Angiò A., Falessi D., Lomartire A., Pesce G., Scarrone S. (2008) Does an intentional architecture pay off? A controlled experiment. Technical Report 09.77, University of Rome TorVergata.

Carver J., Jaccheri L., Morasca S., Shull F. (2003) Issues in using students in empirical studies in software engineering education. Proceedings of the 9th International Symposium on Software Metrics. IEEE Computer Society.

Chen L., Ali Babar M., Cawley C. (2009) Evaluation of variability management approaches: A systematic review. 13th International Conference on Evaluation and Assessment in Software Engineering.

Clements P., Bachmann F., Bass L., Garlan D., Ivers J., Little R., Nord R., Stafford J. (2002a) Documenting software architectures: Views and beyond. Addison-Wesley. Boston.

Clements P., Kazman R., Klein M. (2002b) Evaluating software architecture: Methods and case studies. Addison-Wesley. Boston.

De Marco T. (1986) Controlling software projects. P Hall. New York.

Dehnadi S., Bornat R. (2006) The camel has two humps. http://www.cs.mdx.ac.uk/research/PhDArea/saeed.

Desouza K., Dingsøyr T., Awazu Y. (2005) Experiences with conducting project postmortems: Reports vs. Stories and practitioner perspective. Proceedings of the Proceedings of the 38th Annual Hawaii International Conference on System Sciences (HICSS'05) - Track 8 - Volume 08. IEEE Computer Society.

Dobrica L., Niemelä E. (2002) A survey on software architecture analysis methods. IEEE Transactions on Software Engineering 28 (7), 638-653.

Dybå T., Kitchenham B., Jørgensen M. (2005) Evidence-based software engineering for practitioners. IEEE Software 22 (1), 58-65.

Dybå T., Dingsøyr T. (2008) Empirical studies of agile software development: A systematic review. Information & Software Technology 50 (9-10), 833-859.

Dzidek W., Arisholm E., Briand L. (2008) A realistic empirical evaluation of the costs and benefits of uml in software maintenance. IEEE Trans. Softw. Eng. 34 (3), 407-432.

Easterbrook S., Singer J., Storey M.-A., Damian D. (2008) Selecting empirical methods for software engineering research in. F Shull, J Singer and D Sjøberg, Guide to advanced empirical software engineering Springer.

Eden A., Kazman R. (2003) Architecture, design, implementation. Proceedings of the 25th International Conference on Software Engineering. Portland, Oregon. IEEE Computer Society.

Eeles P. (2005) Capturing architectural requirements. IBM Rational developer works. Available at:http://www.ibm.com/developerworks/rational/library/4706.html.

Eguiluz H. R., Barbacci M. R. (2003) Interactions among techniques addressing quality attributes.

Ellis G., Dix A. (2006) An explorative analysis of user evaluation studies in information visualisation. Proceedings of the 2006 AVI workshop on BEyond time and errors: novel evaluation methods for information visualization. Venice, Italy. ACM.

Engstrom E., Skoglund M., Runeson P. (2008) Empirical evaluations of regression test selection techniques: A systematic review. Proceedings of the Second ACM-IEEE international symposium on Empirical software engineering and measurement. Kaiserslautern, Germany. ACM.

Erdogmus H. (2008) Must software research stand divided? IEEE Software 25 (5), 4-6.

Falessi D., Becker M. (2006) Documenting design decisions: A framework and its analysis in the ambient intelligence domain. BelAmI-Report 005.06/E, Fraunhofer IESE.

Falessi D., Cantone G., Becker M. (2006) Documenting design decision rationale to improve individual and team design decision making: An experimental evaluation. Proceedings of the 5th ACM/IEEE International Symposium on Empirical Software Engineering. Rio de Janeiro, Brazil.

Falessi D., Kruchten P., Cantone G. (2007) Issues in applying empirical software engineering to software architecture. First European Conference on Software Architecture. Aranjunez, Spain. Springer.

Falessi D., Cantone G., Kruchten P. (2008a) Value-based design decision rationale documentation: Principles and empirical feasibility study. Proceeding of the Seventh Working IEEE / IFIP Conference on Software Architecture (WICSA 2008). Vancouver, Canada. IEEE Computer Society.

Falessi D., Capilla R., Cantone G. (2008b) A value-based approach for documenting design decisions rationale: A replicated experiment. Proceedings of the 3rd international workshop on Sharing and reusing architectural knowledge. Leipzig, Germany. ACM.

Glass R. (1994) The software-research crisis. IEEE Software 11 (6), 42-47.

Falessi, D., Ali Babar, M., Cantone, G., Kruchten, P., Empirically Assessing Software Architecture Research: Challenges and Lessons Learned, Empirical Software Engineering Journal, 15(3): pp. 250-276, 2010.

Glass R. (2008) Negative productivity and what to do about it. IEEE Software 25 (5), 96.

Hannay J., Jørgensen M. (2008) The role of deliberate artificial design elements in software engineering experiments. IEEE Transactions on Software Engineering 34 (2), 242-259.

Harrison W. (1998) Sharing the wealth: Accumulating and sharing lessons learned in empirical software engineering research. Empirical Software Engineering 3 (1), 7-8.

Host M., Regnell B., Wohlin C. (2000) Using students as subjects; a comparative study of students and professionals in lead-time impact assessment. Empirical Software Engineering Journal 5 (3), 201-214.

Houdek F. (2003) External experiments: A workable paradigm for collaboration between industry and academia. in. Lecture notes on empirical software engineering World Scientific Publishing Co., Inc.

ISO/IEC 42010 I. P. (2008) Systems and software engineering — architecture description.

Jansen A., Bosch J. (2005) Software architecture as a set of architectural design decisions. 5th Working IEEE/IFIP Conference on Software Architecture (WICSA 5). Pittsburgh. IEEE CS.

Jeffery R., Scott L. (2002) Has twenty-five years of empirical software engineering made a difference? Proceedings of the Ninth Asia-Pacific Software Engineering Conference. IEEE Computer Society.

Ji J., Li J., Conradi R., Liu C., Ma J., Chen W. (2008) Some lessons learned in conducting software engineering surveys in china. Proceedings of the Second ACM-IEEE international symposium on Empirical software engineering and measurement. Kaiserslautern, Germany. ACM.

Jones C. (1994) Assessment and control of software risks. P Hall.

Juristo N., Moreno A. M. (2006) Basics of software engineering experimentation Springer.

Kazman R., Klein M., Clements P. (2004) Atam: Method for architecture evaluation.

Kitchenham B. (2004) Procedures for performing systematic reviews. Joint Technical Report, Keele University TR/SE-0401 and NICTA 0400011T.1.

Kitchenham B., Dybå T., Jørgensen M. (2004) Evidence-based software engineering. Proceedings of the 26th International Conference on Software Engineering. IEEE Computer Society.

Kitchenham B. (2008) The role of replications in empirical software engineering--a word of warning. Journal of Empirical Software Engineering 13 (2), 219-221.

Kitchenham B. A. (1996) 1996. Evaluating software engineering methods and tool part 1: The evaluation context and evaluation methods.

Kruchten P. (2001) Common misconceptions about software architecture. The Rational Edge.

Kruchten P. (2003) The rational unified process: An introduction 3rd Addison-Wesley Professional.

Kruchten P. (2004) An ontology of architectural design decisions in software intensive systems. Proceedings of the 2nd Groningen Workshop on Software Variability.

Kruchten P., Obbink H., Stafford J. (2006) The past, present and future for software architecture. IEEE Software 23 (2), 2-10.

Laitenberger O., Rombach D. (2003) (quasi-)experimental studies in industrial settings. in. Lecture notes on empirical software engineering World Scientific Publishing Co., Inc.

Liu V., Gorton I., Fekete A. (2005) Design-level performance prediction of component-based applications. IEEE Transactions on Software Engineering 31 (11), 928-941.

Lung J., Aranda J., Easterbrook S., Wilson G. (2008) On the difficulty of replicating human subjects studies in software engineering. Proceedings of the 30th international conference on Software engineering. Leipzig, Germany. ACM.

Maranzano J. F., Rozsypal S. A., Zimmerman G. H., Warnken G. W., Wirth P. E., Weiss D. M. (2005) Architecture reviews: Practice and experience. IEEE Software 22 (2), 34-43.

Murphy G., Walker R., Baniassad E. (1999) Evaluating emerging software development technologies: Lessons learned from assessing aspect-oriented programming. IEEE Transaction on Software Engineering 25 (4), 438-455.

Oates B. J. (2003) Widening the scope of evidence gathering in software engineering. Proceedings of the Eleventh Annual International Workshop on Software Technology and Engineering Practice. IEEE Computer Society.

Obbink H., Kruchten P., Kozaczynski W., Hilliard R., Ran A., Postema H., Lutz D., Kazman R., Tracz W., Kahane E. (2002) Report on software architecture review and assessment (sara), version 1.0. At http://philippe.Kruchten.Com/architecture/sarav1.Pdf.

Perry D. E., Porter A., A., Votta L., G. (2000) Empirical studies of software engineering: A roadmap. Proceedings of the Conference on The Future of Software Engineering. Limerick, Ireland. ACM.

Potts C. (1993) Software-engineering research revisited. IEEE Software 10 (5), 19-28.

Falessi, D., Ali Babar, M., Cantone, G., Kruchten, P., Empirically Assessing Software Architecture Research: Challenges and Lessons Learned, Empirical Software Engineering Journal, 15(3): pp. 250-276, 2010.

Prechelt L. (2007) Optimizing return-on-investment (roi) for empirical software engineering studies working group results. in. Empirical software engineering issues. Critical assessment and future directions.

Punter T., Ciolkowski M., Freimut B., John I. (2003) Conducting on-line surveys in software engineering. Proceedings of the 2003 International Symposium on Empirical Software Engineering. IEEE Computer Society.

Reeves J. W. (1992) What is software design? C++ Journal 2 (2).

SEI (2007) Published software architecture definitions. http://www.sei.cmu.edu/architecture/published_definitions.html.

Shaw M., Garlan D. (1996) Software architecture: Perspectives on an emerging discipline. Prentice-Hall. Upper Saddle River, NJ.

Shaw M., Clements P. (2006) The golden age of software architecture. IEEE Software 23 (2), 31-39.

Shull F., Basili V., Carver J., Maldonado J., Travassos G., Mendon M., Fabbri S. (2002) Replicating software engineering experiments: Addressing the tacit knowledge problem. Proceedings of the 2002 International Symposium on Empirical Software Engineering. IEEE Computer Society.

Shull F., Carver J., Hochstein L., Basili V. (2005) Empirical study design in the area of high-performance computing (hpc). International Symposium on Empirical Software Engineering, 2005.

Shull F., Seaman C., Zelkowitz M. (2006) Victor r. Basili's contributions to software quality. IEEE Software 23 (1), 16-18.

Simon H. (1996) The sciences of the artificial 3The MIT Press. Cambridge, Mass.

Sjøberg D., Anda B., Arisholm E., Dybå T., Jørgensen M., Karahasanović A., Vokác M. (2003) Challenges and recommendations when increasing the realism of controlled software engineering experiments. in. Empirical methods and studies in software engineering.

Sjøberg D. I. K., Arisholm E., Jørgensen M. (2001) Conducting experiments on software evolution. Proceedings of the 4th International Workshop on Principles of Software Evolution. Vienna, Austria. ACM.

Sjøberg D. I. K., Anda B., Arisholm E., Tore D., Jørgensen M., Karahasanovic A., Koren E., Marek V. (2002) Conducting realistic experiments in software engineering. Proceedings of the 2002 International Symposium on Empirical Software Engineering. IEEE Computer Society.

Sjøberg D. I. K., Dybå T., Jørgensen M. (2007) The future of empirical methods in software engineering research. Future of Software Engineering, Proceedings of the 29th International Conference on Software Engineering (ICSE 2007). Minneapolis, USA. IEEE Computer Society.

Sjøberg D. I. K., Dybå T., Anda B. C. D., Hannay J. E. (2008) Building theories in software engineering. in. Guide to advanced empirical software engineering.

Smolander K. (2002) Four metaphors of architecture in software organizations: Finding out the meaning of architecture in practice. International Symposium on Empirical Software Engineering (ISESE 2002). Nara, Japan.

Staples M., Niazi M. (2007) Experiences using systematic review guidelines. Journal of Systems and Software 80 (9), 1425-1437.

Svahnberg M., Aurum A., Wohlin C. (2008) Using students as subjects - an empirical evaluation. Proceedings of the Second ACM-IEEE International Symposium on Empirical Software Engineering and Measurement. Kaiserslautern, Germany. ACM.

Tang A., Ali Babar M., Gorton I., Han J. (2007) A survey of architecture design rationale. Journal of Systems & Software 79 (12), 1792-1804

Tonella P., Torchiano M., Bois B., Syst T. (2007) Empirical studies in reverse engineering: State of the art and future trends. Empirical Software Engineering 12 (5), 551-571.

Tyree J., Akerman A. (2005) Architecture decisions: Demystifying architecture. IEEE Software 22 (2), 19-27.

Vegas S., Basili V. (2005) A characterisation schema for software testing techniques. Empirical Software Engineering 10, 437-466.

Vegas S., Juristo N., Moreno A., Solari M., Letelier P. (2006) Analysis of the influence of communication between researchers on experiment replication. Proceedings of the 2006 ACM/IEEE international symposium on Empirical software engineering. Rio de Janeiro, Brazil. ACM.

Vokac M., Tichy W., Sjoberg D., Arisholm E., Aldrin M. (2004) A controlled experiment comparing the maintainability of programs designed with and without design patterns; a

Falessi, D., Ali Babar, M., Cantone, G., Kruchten, P., Empirically Assessing Software Architecture Research: Challenges and Lessons Learned, Empirical Software Engineering Journal, 15(3): pp. 250-276, 2010.

        replication in a real programming environment. Empirical Software Engineering: an International Journal 9 (3), 149-195.

Williams L., Upchurch R. L. (2001) In support of student pair-programming. Proceedings of the 32nd SIGCSE Technical Symposium on Computer Science Education. Charlotte, North Carolina, United States. ACM Press.

Wohlin C., Runeson P., Höst M., Ohlsson M. C., Regnell B., Wesslen A. (2000) Experimentation in software engineering: An introduction. Springer.

Zelkowitz M. V., Wallace D. R. (1998) Experimental models for validating technology. Computer 31 (5), 23-31.